\documentclass[preprint, 11pt, groupedaddress, superscriptaddress, amsfonts, amssymb]{revtex4}
\usepackage{color}
\usepackage{graphicx}
\usepackage[utf8]{inputenc}
\usepackage[T1]{fontenc}
\usepackage[right]{eurosym}
\usepackage{calc}
\usepackage{xspace}
\usepackage{booktabs}
\usepackage{natbib}
   
\def€{\EUR}

\def\fm#1{\ifmmode #1 \else $#1$\fi}
\def\ket#1{{%
  \ifmmode |\,#1\,\rangle \else $|\,#1\,\rangle$\fi}}
\def\bra#1{{%
  \ifmmode \langle\,#1\,| \else $\langle\,#1\,|$\fi}}
\def\braket#1#2{{%
  \ifmmode \langle\,#1\,|\,#2\,\rangle \else $\langle\,#1\,|\,#2\,\rangle$\fi}}
\def\expect#1{{%
  \ifmmode \langle\,#1\,\rangle \else $\langle\,#1\,\rangle$\fi}}
\def\upket{\fm{\ket{\!\uparrow}}}
\def\downket{\fm{\ket{\!\downarrow}}}

\def\Ca{\fm{\mathrm{Ca}^{+}}\xspace}
\def\Cafz{\fm{^{40}\Ca}\xspace}
\def\Mg{\fm{\mathrm{Mg}^{+}}\xspace}
\def\Mgtf{\fm{^{25}\Mg}\xspace}

\def\dsoh{\fm{{}^2\mathrm{S}_{1/2}}\xspace}
\def\dpoh{\fm{{}^2\mathrm{P}_{1/2}}\xspace}
\def\dpth{\fm{{}^2\mathrm{P}_{3/2}}\xspace}
\def\ddth{\fm{{}^2\mathrm{D}_{3/2}}\xspace}

\def\dsohdpoh{\dsoh\fm{\leftrightarrow}\dpoh\xspace}

\begin{document}
\title{Precision spectroscopy by photon-recoil signal amplification}.
\author{Yong Wan}
\thanks{These authors contributed equally.}
\affiliation{QUEST  Institut, Physikalisch-Technische Bundesanstalt, 38116 Braunschweig, Germany}
%\homepage[]{www.quantummetrology.de/quest/eqm}
\author{Florian Gebert}  
\thanks{These authors contributed equally.}
\affiliation{QUEST  Institut, Physikalisch-Technische Bundesanstalt, 38116 Braunschweig, Germany}
\author{Jannes B. Wübbena}
\affiliation{QUEST  Institut, Physikalisch-Technische Bundesanstalt, 38116 Braunschweig, Germany}
\author{Nils Scharnhorst}
\affiliation{QUEST  Institut, Physikalisch-Technische Bundesanstalt, 38116 Braunschweig, Germany}
\author{Sana Amairi}
\affiliation{QUEST  Institut, Physikalisch-Technische Bundesanstalt, 38116 Braunschweig, Germany}
\author{Ian D. Leroux}
\affiliation{QUEST  Institut, Physikalisch-Technische Bundesanstalt, 38116 Braunschweig, Germany}
\author{Börge Hemmerling}
\thanks{present address:  Department of Physics, Harvard University, Cambridge, Massachusetts 02138, USA}
\affiliation{QUEST  Institut, Physikalisch-Technische Bundesanstalt, 38116 Braunschweig, Germany}
\author{Niels Lörch}
\affiliation{Institut für Theoretische Physik und Institut für Gravitationsphysik (Albert Einstein Institut), Leibniz Universität Hannover, 30167 Hannover, Germany}
\author{Klemens Hammerer}
\affiliation{Institut für Theoretische Physik und Institut für Gravitationsphysik (Albert Einstein Institut), Leibniz Universität Hannover, 30167 Hannover, Germany}
\author{Piet O. Schmidt}
\email{Piet.Schmidt@quantummetrology.de}
\affiliation{QUEST  Institut, Physikalisch-Technische Bundesanstalt, 38116 Braunschweig, Germany}
%\homepage[]{www.quantummetrology.de/quest/eqm}
\affiliation{Institut für Quantenoptik, Leibniz Universität Hannover, 30167 Hannover, Germany}

\date{\today}

\begin{abstract}
Precision spectroscopy of atomic and molecular ions offers a window to new
physics, but is typically limited to species with a cycling transition for laser
cooling and detection. Quantum logic spectroscopy has overcome this limitation
for species with long-lived excited states. Here we extend quantum logic
spectroscopy to fast, dipole-allowed transitions and apply it to perform an
absolute frequency measurement. We detect the absorption of photons by the
spectroscopically investigated ion through the photon recoil imparted on a
co-trapped ion of a different species, on which we can perform efficient quantum
logic detection techniques. This amplifies the recoil signal from a few absorbed
photons to thousands of fluorescence photons. We resolve the line center of a
dipole-allowed transition in \Cafz to $1/300$ of its observed linewidth,
rendering this measurement one of the most accurate of a broad transition. The
simplicity and versatility of this approach enables spectroscopy of many
previously inaccessible species.
\end{abstract}

\maketitle

\section{Introduction}
Precision optical spectroscopy of broad transitions provides information on the structure of molecules \cite{laane_frontiers_2011}, it allows tests of quantum electrodynamics \cite{karshenboim_hydrogen_2001}, and, through comparison with astrophysical data, probes for a possible variation of fundamental constants over cosmological scales \cite{webb_indications_2011, rahmani_constraining_2012}. Nuclear properties are revealed through isotope shift measurements \cite{nortershauser_isotope_1998, palmer_laser_1999, batteiger_precision_2009, lee_high_2013, blaum_precision_2013}, or absolute frequency measurements \cite{pohl_size_2010, antognini_proton_2013}.
Trapped ions are particularly well suited for such high precision experiments. The ions are stored in an almost field-free environment and can be laser-cooled to eliminate Doppler shifts. These features have enabled record accuracies in optical clocks \cite{rosenband_frequency_2008, chou_frequency_2010, huntemann_high-accuracy_2012, madej_<sup>88</sup>sr<sup>+</sup>_2012}. 
For long-lived excited states such as in atoms with clock transitions, the electron shelving technique amplifies the signal from a single absorbed photon by scattering many photons on a closed transition through selective optical coupling of one of the two spectroscopy states to a third electronic level \cite{dehmelt_shelved_1975}. The invention of quantum logic spectroscopy (QLS) \cite{schmidt_spectroscopy_2005, rosenband_frequency_2008} removed the need to detect the signal on the spectroscopically investigated ion (spectroscopy ion) by transferring the internal state information through a series of laser pulses to the co-trapped, so-called logic ion where the signal is observed via the electron-shelving technique. However, this original implementation of QLS requires long-lived spectroscopy states to implement the transfer sequence. 
For transitions with a short-lived excited state, spectroscopy of trapped ions is typically implemented through detection of scattered photons in laser induced fluorescence (LIF) \cite{drullinger_high-resolution_1980, nakamura_laser_2006, batteiger_precision_2009, wolf_frequency_2008, takamine_isotope_2009, herrera-sancho_energy_2013} or detection of absorbed photons in laser absorption spectroscopy (LAS) \cite{wineland_absorption_1987}.  Neither of the two techniques reaches the fundamental quantum projection noise limit as in the electron shelving technique \cite{itano_quantum_1993} due to low light collection efficiency in LIF and small atom-light coupling in LAS. In a variation of absorption spectroscopy, the detuning-dependent effect of spontaneous light forces on the motional state of trapped ions is employed to obtain a spectroscopy signal. This has been investigated by observing macroscopic heating on a Doppler-cooled two-ion crystal  \cite{clark_detection_2010}, and through a measurement of motional phase change due to photon scattering \cite{hempel_entanglement-enhanced_2013}.

Here, we present and demonstrate an extension of quantum logic spectroscopy
which combines the advantages of these techniques and provides a highly
sensitive, quantum projection noise-limited signal for the spectroscopy of broad
transitions. It is based on detection of momentum kicks from a few absorbed
photons near the resonance of a single spectroscopy ion through a co-trapped
logic ion. We demonstrate photon recoil spectroscopy (PRS) on the \dsohdpoh transition in a \Cafz ion, resolving the
line by a factor of $1/300$ of its observed linewidth of $\sim 34$~MHz and achieve a
statistics-limited fractional instability of $\sigma_y(\tau)=5.1\times
10^{-9}\sqrt{\mathrm{s}/\tau}$, corresponding to 100 absorbed photons for a resolution of 7.2~MHz. This proof-of-principle experiment demonstrates the potential of PRS and its applicability to different species.

\section{Results}
\subsection{Spectroscopy Scheme}
\begin{figure}
	\includegraphics[width=12cm]{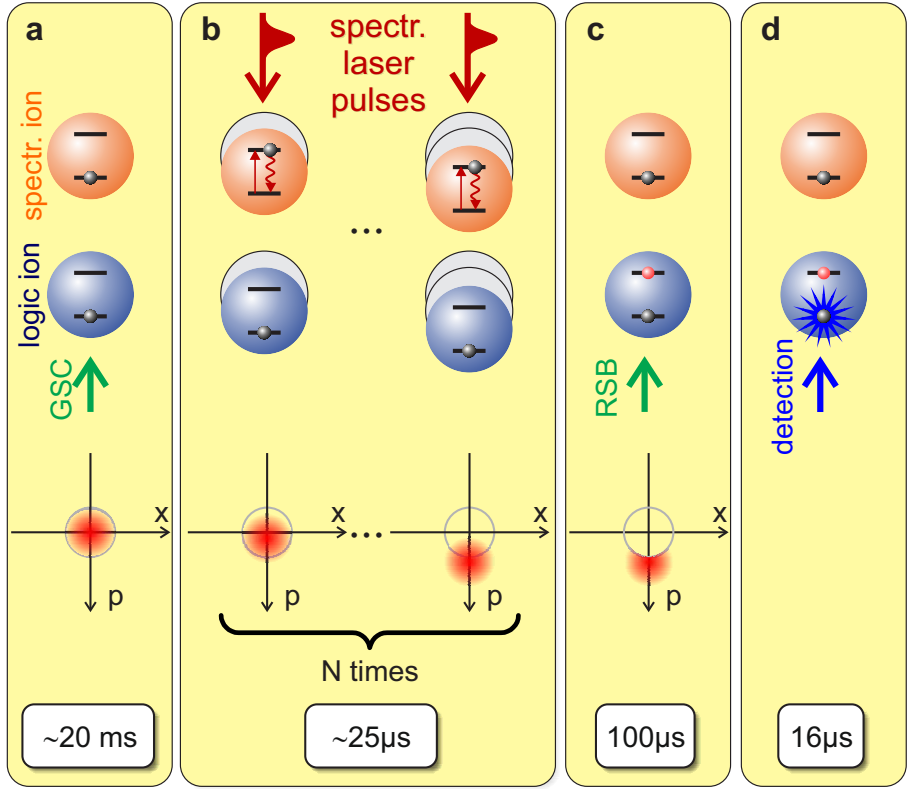}
	\caption{	\label{fig:principle}
	Principle of photon recoil spectroscopy. The top part of the figure shows the internal and motional excitation of the two-ion crystal (spectroscopy ion: top, logic ion: bottom). The center part shows the Wigner phase-space representation of the excited normal mode. Typical timescales for the experimental implementation are shown in the bottom part. \textbf{a} Initially, the two-ion crystal is prepared in the ground state of motion through ground state cooling (GSC) on the logic ion. \textbf{b} Absorption of photons from the pulsed spectroscopy laser results in excitation of motion due to photon-induced recoil. Motional displacements are only in momentum, since the pulses are synchronized with the oscillation period of the normal mode. \textbf{c} The motionally excited part of the wavefunction containing Fock states \ket{n>0} (represented by the red moon-shaped disc in the phase-space representation) that have no overlap with the ground state \ket{0} is converted into an internal state excitation of the logic ion by a transfer pulse on the red-sideband (RSB). \textbf{d} Electron-shelving detection of the internal state of the logic ion provides the spectroscopy signal.}
\end{figure}
The principle of PRS is shown in Fig.~\ref{fig:principle}. Starting from the axial motional ground state of the two-ion crystal  (Fig.~\ref{fig:principle}a), momentum transfer from photon absorption on the spectroscopy ion results in motional excitation (Fig.~\ref{fig:principle}b). Owing to the strong Coulomb interaction, this excitation is shared by the logic ion where it is converted into a long-lived internal excitation by applying a so-called red-sideband laser pulse  which changes the internal state conditioned on the presence of motional excitation (Fig.~\ref{fig:principle}c) \cite{wineland_experimental_1998, leibfried_quantum_2003, haffner_quantum_2008}.
Subsequent electron-shelving detection on the logic ion amplifies the signal of a few absorbed photons on the spectroscopy ion to thousands of scattered photons on the logic ion, providing quantum projection noise-limited detection of motional excitation. This signal is background-free, since spectroscopy and detection are separated in time.

During interaction with spectroscopy pulses, absorption of a photon displaces the motional state in momentum by $\hbar k$, where $k$ is the wavenumber of the spectroscopy light. Efficient excitation of a chosen normal mode is accomplished through application of short (compared to the oscillation period $T_\mathrm{m}$ in the trap) spectroscopy laser pulses synchronized to the respective normal mode frequency $\omega_\mathrm{m}=2\pi/T_\mathrm{m}$ \cite{drewsen_nondestructive_2004, lin_resonant_2013, hempel_entanglement-enhanced_2013}.  The momentum displacement of this mode upon absorption of a single photon is given by the Lamb-Dicke factor $\eta$ (in units of the zero-point momentum fluctuations of the normal mode) \cite{wubbena_sympathetic_2012}. This motional excitation is shared by the logic ion. It is then converted into an electronic excitation of the logic ion via a transfer pulse on the red sideband. In a simple model, the excitation signal on the logic ion after this pulse is proportional to the probability of leaving the motional ground state, which is approximately given by 
\begin{equation}\label{eq:pe}
p_\mathrm{e}\approx 1-\exp(-N_\mathrm{a}^2\eta^2),
\end{equation}
where $N_\mathrm{a}$ is the number of absorbed photons. Choosing $p_\mathrm{e}=0.5$ yields a SNR of 1 in a single experiment, the best allowed by quantum projection noise \cite{itano_quantum_1993}. For typical Lamb-Dicke parameters of $0.1$ this requires absorption of around 10 photons.

An analytical treatment of the observed signal based on a Fokker-Planck equation qualitatively confirms the favorable scaling with the number of absorbed photons and can be found in Supplementary Note 2. 

\begin{figure}
	\includegraphics[width=14cm]{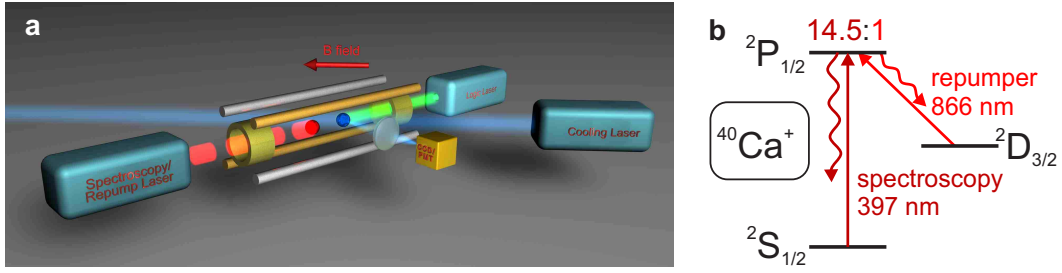}
	\caption{\label{fig:setup}Experimental system for photon recoil spectroscopy. \textbf{a}  Sketch of the experimental setup showing the linear ion trap with spectroscopy ion (red) and logic ion (blue), the laser beam and magnetic field directions, and the detection system for the logic ion. Doppler cooling, detection, and sideband manipulation using stimulated Raman transitions are performed on the \Mgtf logic ion with light around 280~nm. The two Raman beams are depicted by a single beam corresponding to their relative wave vector $\Delta k$. \textbf{b} Relevant level scheme for the \Cafz spectroscopy ion. Spectroscopy and repumping on \Cafz require lasers at wavelengths of 397~nm and 866~nm, respectively. The excited \dpoh-state can decay back to the \dsoh or into the long-lived metastable \ddth-state  with a branching ratio of 14.5:1.}
\end{figure}
\subsection{Implementation}
We have implemented photon recoil spectroscopy using \Mgtf as the logic ion to perform an absolute frequency measurement of the  \dsohdpoh transition in \Cafz with a natural linewidth of 21.6~MHz. % \cite{arora_blackbody-radiation_2007}. 
A sketch of the experimental setup together with the level scheme of the
spectroscopy ion is shown in Fig.~\ref{fig:setup}. Details of the system have
been described previously \cite{hemmerling_single_2011, hemmerling_novel_2012}
and are summarized in the Methods section.

\begin{figure}
	\includegraphics[width=14cm]{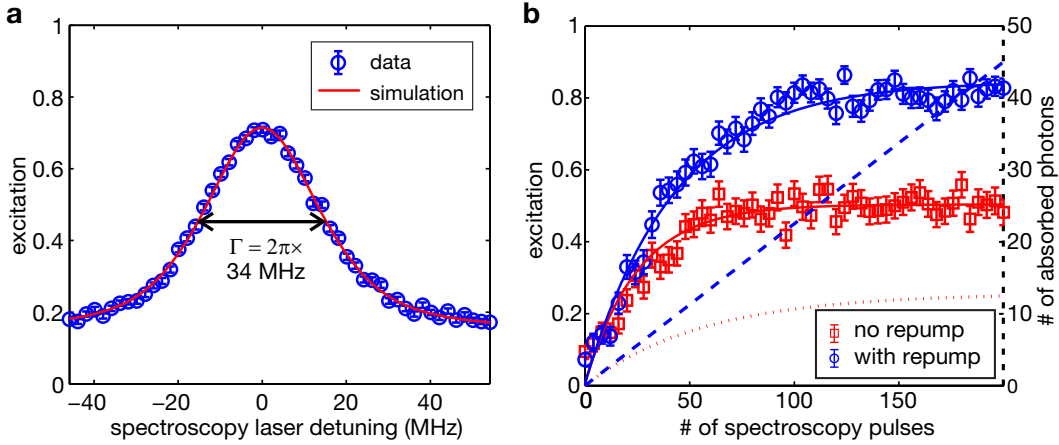}
	\caption{\label{fig:results}Photon recoil spectroscopy results. \textbf{a} Frequency scan across the resonance line. The red line is the result of a numerical simulation using calibrated parameters and has been adjusted in amplitude by 0.93 and shifted by 0.15 to account for experimental imperfections (see text for details). \textbf{b} On-resonance excitation signal as a function of the number of spectroscopy pulses with (blue circles) and without (red squares) repumper. The solid lines (left axis) are numerical simulations from which the Rabi frequency $\Omega$ of the spectroscopy laser is extracted. The dashed and dotted lines (right axis) are derived from the same simulations and give the corresponding  integrated number of absorbed photons with and without repumper, respectively. All error bars correspond to $1\sigma$ statistical uncertainty.}
\end{figure}
\subsection{Resonance line and sensitivity}
Fig.~\ref{fig:results}a shows the average of 12 scans across the resonance of the \dsohdpoh transition in \Cafz using photon recoil spectroscopy as described in Fig.~\ref{fig:principle}. Typically 70 spectroscopy pulses per cycle are applied and a cycle is repeated 250 times  for each frequency point. After each spectroscopy pulse, a repump pulse is applied to clear out the metastable \ddth state.
The observed linewidth of the transition is $\sim 34$~MHz,  compatible with Zeeman broadening in the applied bias magnetic field of $5.84(1)\times 10^{-4}$~T and Fourier broadening due to the 50~ns short spectroscopy pulses. An additional line broadening due to saturation of the excitation signal for large motional displacements (ground state depletion) is avoided by choosing a small number of absorbed photons. 
The line is a numerical simulation of the spectroscopy signal using a full
master equation model (see Supplementary Note 3) and corrected for experimental
imperfections, such as a reduced signal amplitude (typically $\geq 0.85$)  from limited sideband transfer efficiency and a signal offset of typically $\leq 0.15$ due to incomplete ground state cooling.
The photon sensitivity of the signal was determined by taking advantage of the excited \dpoh state branching ratio (see Fig.~\ref{fig:setup}b). For this, the transition was excited on resonance and the signal amplitudes with and without repumping pulses on \Cafz were compared. Fig.~\ref{fig:results}b shows the result as a function of the number of spectroscopy pulses. The Rabi frequency of the spectroscopy laser and the signal per scattered photon is extracted from a numerical simulation of the spectroscopy signal, which takes the excited state branching ratio of 14.5:1 \cite{ramm_precision_2013} into account, corresponding to 15.5 absorbed photons in the absence of any repumping.
An excitation of $p_\mathrm{e}=0.5$ corresponding to $\mathrm{SNR}=1$ in a
single experiment is achieved by absorbing 9.5(1.2) photons, in qualitative
agreement with equation~\ref{eq:pe} for our experimental parameters and
confirmed by the analytical model (Supplementary Note 2). This high sensitivity
even for a small Lamb-Dicke factor of $\eta=0.108$ is a consequence of the exponential scaling with the square of the number of absorbed photons (see equation~\ref{eq:pe}).

\begin{figure}
	\includegraphics[width=14cm]{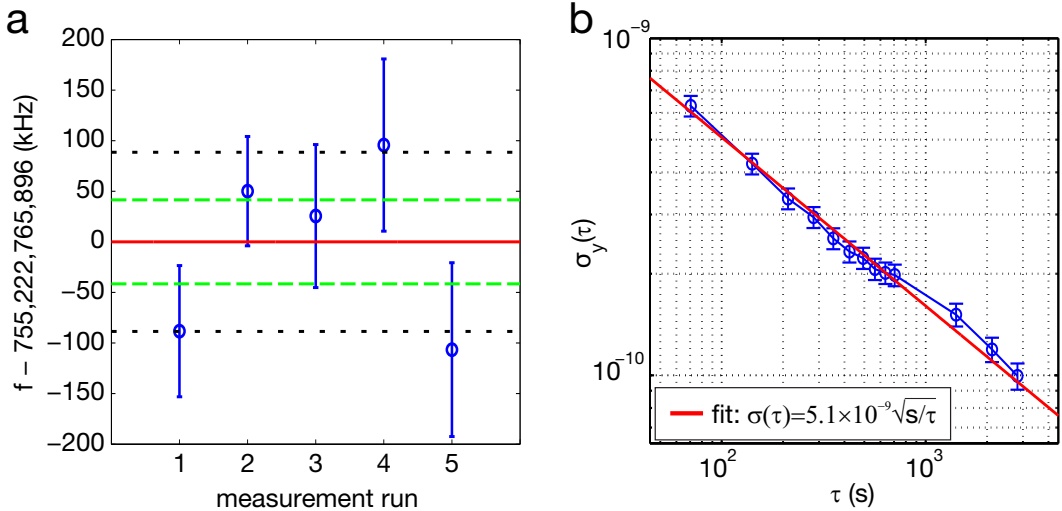}
	\caption{\label{fig:freqmeas}Absolute frequency measurement results. \textbf{a}
	Compilation of absolute frequency measurements performed using the two-point
	sampling technique.
	The dotted black and dashed green lines give the 68.3~\%
	prediction bounds for the measurement and the mean, respectively. \textbf{b} Allan deviation of the combined data set. A fit gives an instability of  $\sigma_y(\tau)=5.1\times 10^{-9}\sqrt{(\mathrm{s}/\tau)}$. The error bars indicate the standard deviation of the
	mean.}
\end{figure}
\subsection{Absolute frequency measurement}
The main result of this work is shown in Fig.~\ref{fig:freqmeas}, where we demonstrate that PRS enables precision spectroscopy at an unprecedented level.
It shows the results of a frequency measurement of  the \dsohdpoh transition in \Cafz, which we have implemented using a two-point sampling technique \cite{itano_quantum_1993}. For this, we set the probe laser frequency close to the center of the atomic resonance and probe the excitation probability for constant frequency steps towards the left and right of the resonance center. The center frequency is determined from the average of the excitation probabilities $p_\mathrm{l,r}$ on each side, multiplied with the frequency discriminant given by the experimentally determined slope of the resonance curve.
These steps are repeated until the desired frequency uncertainty is achieved. The frequency measurements shown in Fig.~\ref{fig:freqmeas}a have been performed on 5 days within 11 weeks.
Fig.~\ref{fig:freqmeas}b shows the relative frequency uncertainty as a function
of averaging time $\tau$ in form of an overlapping Allan deviation
$\sigma_y(\tau)$ \cite{riehle_frequency_2004} of the concatenated five
measurement runs. The observed instability $\sigma_y(\tau)=5.1\times
10^{-9}\sqrt{\mathrm{s}/\tau}$ (red line) is in agreement with the expected
Allan deviation of $5.8(9)\times 10^{-9}$ given the measured SNR for a single
ion using the true measurement time, including dead time from ground state
cooling and detection. This confirms that the measurement is statistics-limited.
The required averaging time can be reduced by at least a factor of ten through
speed-optimized ground state cooling \cite{lin_sympathetic_2013}. For chosen
excitation probabilities $p_\mathrm{l,r}$, the mean number of scattered photons
for each probe cycle can be determined from the calibration shown in
Fig.~\ref{fig:results}b. This allows us to express the relative frequency
uncertainty as a function of the number $N_\mathrm{a}$ of absorbed photons,
which is given by $\sigma_y(N_\mathrm{a})=9.5\times
10^{-8}/\sqrt{N_\mathrm{a}}$, corresponding to an uncertainty of 7.2~MHz for 100
absorbed photons.

\begin{table}
\begin{tabular}{|l|c|}
\hline effect (origin) & shift (kHz) \\ 
\hline Zeeman (magnetic field) & $-8\pm 59$ \\ 
\hline AC Stark (spectroscopy laser) & $60\pm 44$ \\ 
\hline spectroscopy pulse envelope shift (bandwidth) & $0\pm 16$\\
\hline lineshape (detection scheme) & $151\pm 20$ \\ 
\hline statistics &  $0\pm 42$\\ 
\hline 
\hline total & $203\pm 88$\\
\hline  
\end{tabular} 
\caption{\label{tab:systematics}Uncertainty evaluation of the absolute frequency
measurement. The given uncertainties correspond to $1\sigma$ errors. Details can
be found in the text and the Supplementary Note 1 and 3. The spectroscopy pulse
envelope shift is on the order of $\pm 50$~kHz and is calibrated individually for each measurement.}
\end{table}

Table~\ref{tab:systematics} shows a summary of the evaluated systematic effects
for the absolute frequency measurement. Details can be found in the
Supplementary Note 1 and 3. Lineshape asymmetries from an excitation imbalance
of Zeeman-shifted line components are strongly suppressed by symmetrical driving of the transition by linearly polarized light.
Off-resonant coupling of the spectroscopy laser to other states can cause frequency-dependent AC Stark shifts. We have evaluated both shifts by measuring the line center frequency as a function of the magnetic field and spectroscopy laser intensity. As expected, the results are compatible with vanishing shifts, limited by the statistical uncertainty of the measurement. 
Shifts arising from micromotion-induced AC Stark and second order Doppler effects, as well as lineshape and frequency uncertainties arising from the spectroscopy laser stabilization have been estimated to be well below 1~kHz.
The spectral bandwidth of the 125~ns spectroscopy pulses used in the frequency measurements, together with the limited bandwidth of the switching device (an acousto-optical modulator), can cause a shift in the center frequency of the spectroscopy pulses on the order of $\pm 50$~kHz. We calibrate this shift for each experimental run to within 16~kHz.

Probe-frequency-dependent Doppler heating and cooling effects during excitation result in an asymmetric lineshape of the motional excitation signal.
Starting from the motional ground state in our system, we have a precise knowledge of the Doppler-induced shifts through the excitation signal. This allows an accurate analytical and numerical calculation of the shift.
From a simple model taking into account Doppler induced (anti)damping and associated narrowing (spread) of the wave packet in momentum space, this frequency shift is given by
\begin{equation}\label{eq:shift}
\Delta \omega=\frac{\eta\omega_\mathrm{m}}{2\sqrt{\log \left(\frac{1}{1-p_\mathrm{l,r}}\right)}}=\frac{\omega_\mathrm{m}}{2N_\mathrm{a}},
\end{equation}
where we have substituted equation~\ref{eq:pe} for $p_\mathrm{l,r}$ in the last step.
The shift increases for smaller excitation probabilities $p_\mathrm{l,r}$ used in the two-point sampling technique, corresponding to a larger detuning of the two probe frequencies from resonance and a smaller number of scattered photons. It is worthwhile noting that within the approximations of the model, this shift is independent of the choice of other parameters, such as spectroscopy laser Rabi frequency or the transition linewidth.
The result from this simple model agrees very well for typical excitation
probabilities between 0.1 and 0.9 with analytical and numerical models including
diffusion and drift in position (Supplementary Note 2 and 3). For smaller excitation
probability, the shift is bounded by the trap frequency.
From conservatively estimated fluctuations of the detection signal of $\pm 0.1$ around $p_\mathrm{l,r}=0.5$, we account  for a shift of $151(20)$~kHz.
The uncertainty of this residual Doppler shift can be significantly reduced, while at the same time improving the photon sensitivity, by lowering the trap frequency during spectroscopy. A similar, but much stronger effect from heating/cooling processes during spectroscopy of Doppler-cooled ions has been reduced by sympathetic cooling \cite{herrmann_frequency_2009} and interleaved cooling/spectroscopy pulses \cite{wolf_frequency_2008}.
 
The absolute frequency averaged over five individual measurements  is $755,222,765,896 (88)$~kHz, of which $78$~kHz are from systematic and $42$~kHz from statistical uncertainty, in agreement with a previous measurement~\cite{wolf_frequency_2008}, but with an accuracy improved by more than an order of magnitude. 
To our knowledge this result represents the most accurate absolute frequency measurement of a broad transition in a trapped ion system, resolving the line center by a factor of more than 300 of the transition's observed linewidth \cite{wolf_frequency_2008, batteiger_precision_2009, herrmann_frequency_2009}. 

\section{Discussion}
The high photon sensitivity and absolute frequency accuracy of PRS together with
its simplicity, requiring only ground state cooling, enables spectroscopy of atomic and molecular species inaccessible with other techniques. The method is particularly well suited for systems with non-closed transitions or intrinsically low fluorescence rates, such as molecular ions \cite{lien_optical_2011, nguyen_challenges_2011, nguyen_prospects_2011}, astro-physically relevant metal ions with a complex level structure \cite{webb_indications_2011, rahmani_constraining_2012, berengut_atomic_2011}, and highly-charged ions \cite{nortershauser_laser_2011}. PRS is also an efficient replacement for the original quantum logic spectroscopy implementation for long-lived excited states, since it does not require the driving of sideband transitions on the spectroscopy ion, thus lowering the requirements on the spectroscopy laser. Combined with the single-photon sensitivity for high photon energies and the background-free detection, PRS enables direct frequency comb \cite{marian_united_2004, diddams_molecular_2007}, XUV \cite{jones_phase-coherent_2005, gohle_frequency_2005}, and  x-ray spectroscopy \cite{epp_soft_2007, bernitt_unexpectedly_2012} which offer only a small number of photons per spectral bandwidth. This will pave the way for spectroscopic applications ranging from chemistry to fundamental questions in atomic, molecular, and nuclear physics.
 
\section{Methods}
\subsection{Experimental methods}
\begin{figure}[b]
	\includegraphics[width=5.5cm]{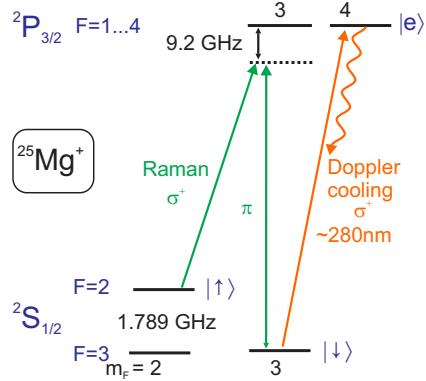}
	\caption{	\label{fig:logic}
	Relevant energy levels and transitions in the magnesium logic ion. Ground-state cooling and coherent manipulation are implemented through Raman beams (detuned by 9.2~GHz from the excited state), which couple the two qubit states $\ket{F=3, m_F=3}=\downket$, $\ket{F=2,m_F=2}=\upket$ using $\pi$- and $\sigma$-polarized light. Laser cooling and internal state discrimination is performed using resonant light with the $\downket\leftrightarrow \ket{F=4, m_F=4}=\ket{e}$ excited state.}
\end{figure}

We trap a two-ion crystal consisting of \Mgtf and \Cafz in a linear ion trap at single-\Mgtf radial and axial trap frequencies of $\sim 5$~MHz and 2.22~MHz,  respectively. Using resolved sideband cooling \cite{hemmerling_single_2011} on \Mgtf, we cool the in-phase (ip) and out-of-phase (op) axial modes with frequencies $f_\mathrm{ip}=1.92$~MHz and $f_\mathrm{op}=3.51$~MHz to their ground state of motion with a typical residual mean occupation of $\bar{n}_\mathrm{ip}\approx 0.07$ and $\bar{n}_\mathrm{op}\approx 0.02$. We have chosen the ip-mode as the spectroscopy mode, since it has the largest Lamb-Dicke parameter of $\eta_\mathrm{Ca^+, ip}=\eta=0.108$ for the calcium ion. The RSB pulse on the logic ion is implemented by a stimulated Raman transition that is red detuned from the $\ket{F=3, m_F=3}=\downket \leftrightarrow \upket=\ket{F=2, m_F=2}$ transition in the $\dsoh$ state (see Fig.~\ref{fig:logic}) by the normal mode frequency. Since the required $\pi$-time for optimum transfer depends on the initial motional state \cite{wineland_experimental_1998}, we employ a stimulated Raman adiabatic passage (STIRAP)-like pulse by linearly ramping down the $\pi$-Raman beam intensity during 100~$\mu$s while ramping up the intensity of the $\sigma$-Raman beam. Although not fully adiabatic, we achieve a more than 80~\% transfer efficiency of all motionally excited states from $\downket$ to $\upket$, independent of their initial state. Internal state discrimination is performed by applying light resonant with the cycling transition $\downket\rightarrow\ket{e}=\ket{\dpth, F=4, m_F=4}$ (see Fig.~\ref{fig:logic}) using the $\pi$-detection technique \cite{hemmerling_novel_2012}.

The 397~nm spectroscopy laser consists of a frequency-doubled extended cavity diode laser. The laser light can be frequency tuned via a double-pass  acousto-optical modulator (AOM). An additional AOM is used to provide short ($50\dots 125$~ns) pulses of spectroscopy light. The intensity-stabilized light is delivered to the atom via a single-mode optical fiber to avoid beam steering errors. The laser is phase-locked to an optical frequency comb which is referenced to one of the hydrogen masers at the German National Metrology Institute (PTB), providing a direct link to the SI second (PTB) with a relative frequency uncertainty of better than $10^{-12}$ for averaging times of around one hour. The spectroscopy and repump light are linearly polarized in a plane spanned by the magnetic field and the trap axis. Under ideal conditions, this results in equal amounts of $\sigma^+$ and $\sigma^-$ polarized light and an additional $\pi$ component, providing a symmetric spectrum, independent of the magnetic field.
%For the frequency measurements, we randomly choose one side and probe the excitation probability $1250$~times at $p_{l,r}\approx 0.5$.

\subsection{Frequency measurements}
For the final frequency measurement, we employ the two point sampling method \cite{itano_quantum_1993}, where the spectroscopy laser probes the excitation signal at the two frequencies ($\nu_\mathrm{l}=\nu_\mathrm{probe} - \Delta \nu_\mathrm{FWHM}/2$, $\nu_\mathrm{r}=\nu_\mathrm{probe} + \Delta \nu_\mathrm{FWHM}/2$), which ideally provides an excitation of half the maximum excitation. In every single scan the experiment is repeated on both frequencies for $5\times 250=1250$ times. To reduce the detection error, we employ the $\pi$-detection method \cite{hemmerling_novel_2012} with threshold at $0.5$ to extract the excitation probability. This detection scheme is based on post-selection, which inherently reduces the effective number of cycles to a typical value of $N^\prime_\mathrm{cycles} = 1039(12)$. This corresponds to a quantum projection noise of $\sigma^\mathrm{(p)} = \sqrt{p(1-p)/N_\mathrm{cycles}} \approx 0.0155$. The difference of excitation at the two probe frequencies $\delta p = p(\nu_\mathrm{l}) - p(\nu_\mathrm{r})=p_\mathrm{l}-p_\mathrm{r}$ is used as an error signal to steer the probe frequency $\nu_\mathrm{probe}$ towards the line center in the following scan.
The uncertainty for the quantity $\delta p$ rescales to $\sigma^\mathrm{(\delta p)} = \sqrt{\sigma^\mathrm{p}_\mathrm{l} + \sigma^\mathrm{p}_\mathrm{r}} \approx 0.0220$.

%To determine the frequency discriminant, we scan the frequency of the spectroscopy laser across a small range of about 4 MHz around $\nu_\mathrm{probe} - \Delta \nu_\mathrm{FWHM}/2$ and $\nu_\mathrm{probe} + \Delta \nu_\mathrm{FWHM}/2$ (Fig. \ref{pic:freq_discriminant}). Linear fits of the two sets of data gives the slopes $(k_\mathrm{1}, k_\mathrm{2})$ of the resonance curve at around $\nu_\mathrm{probe} - \Delta \nu_\mathrm{FWHM}/2$ and $\nu_\mathrm{probe} + \Delta \nu_\mathrm{FWHM}/2$.

The excitation difference $\delta p$ is turned into an absolute frequency $\nu$ by the following relation
\begin{equation}
		\nu = \nu_\mathrm{probe} - D\cdot \delta p,
\end{equation}
where the frequency discriminant $D$ is calibrated for each measurement day experimentally through a measurement of the signal slope around the probe frequencies $\nu_\mathrm{l,r}$.

The final statistical uncertainty for the frequency measurement propagates as
\begin{equation}\label{eq:qpn}
		\sigma_\mathrm{\nu} = D|\delta p| \sqrt{ \left(\frac{\sigma^\mathrm{(\delta p)}}{\delta p}\right)^2 + \left(\frac{\sigma^\mathrm{D}}{D}\right)^2},
\end{equation}
where the first term contains the quantum projection noise and the second one the calibration error for the frequency discriminant.

Concatenating the time-resolved frequency uncertainty from the five measurements shown in Fig.~4a, the overlapping Allan-deviation is derived as shown in Fig. 4b. This gives an inferred instability of $5.1\times 10^{-9}$ for an averaging time of 1~s. This result is in agreement with the quantum projection noise limit of $5.8(9)\times 10^{-9}$ estimated according to
\begin{equation}
		\sigma_\mathrm{\nu}^\mathrm{QPN} = \frac{D\sigma^\mathrm{(\delta p)}}{\nu}\sqrt{T},
\end{equation}
%which is shown as the grey shaded area in Fig.~4b. Here T ($= 70.5$~s) refers to the time for repeating the experimental cycles 1250 times.
where T ($= 70.5$~s) refers to the time for repeating the experimental cycles 1250 times.

The data shown in Fig.~4 is based on a selection of data that has passed statistical tests to check its validity.
A total of seven measurement runs, each performed on a different day, have passed normality tests (Kolmogorov-Smirnov test, Lillifors test, Jarque-Bera test, Shapiro-Wilk test). In a second step, two sample tests (two-sample Kolmogorov-Smirnov test, two-sample t-test) were performed on every combination of two measurements to verify that all the measurements come from the same normal distribution. Two of them did not pass these tests and were excluded from Fig.~4. We suspect that the data was contaminated by drifts in the experimental parameters, such as STIRAP and ground-state cooling efficiency, and calibration of the pulse spectrum.

After excluding the data that fail the distribution tests, the weighted average of all the frequencies measured on a single day is computed according to
\begin{equation}
		\bar{\nu} = \frac{\sum\limits_{i} w^{(i)}\nu^{(i)}}{\sum\limits_{i} w^{(i)}};
\end{equation}
with the weight factor $w^{(i)} = 1/(\sigma^{(i)}_{\nu})^2$. The variance of the mean equals
\begin{equation}
		\sigma_{\bar{\nu}}^2 = \frac{1}{N}\frac{\sum\limits_{i} w^{(i)}(\nu^{(i)} - \bar{\nu})^2}{\sum\limits_{i} w^{(i)}}
\end{equation}
with $N$ as the number of scans on that measurement day.

A weighted fit of the center frequencies from different measurement days with the weights $1/\sigma_\mathrm{\bar{\nu}}^2$ as in Fig. 4a gives the center frequency $\nu = 755,222,765,896$~kHz with a statistical uncertainty of 42~kHz.

\begin{acknowledgments}  
We acknowledge the support of DFG through QUEST and grant SCHM2678/3-1.  YW acknowledges support from IGSM, JBW acknowledges support from HALOSTAR and Studienstiftung des deutschen Volkes, IDL acknowledges support from the Alexander von Humboldt Foundation.  This work was supported by the European Metrology Research Programme (EMRP) in project SIB04. The EMRP is jointly funded by the EMRP participating countries within EURAMET and the European Union. We thank Andreas Bauch for providing the maser signal, Harald Schnatz for help with calibrations, Harald Telle for stimulating discussions, and P. Carstens for technical support.
\end{acknowledgments}
  
\section*{Author Contributions}
The experimental setup was built by B.H., F.G., Y.W., and P.O.S. The spectroscopy laser was set up and characterized by J.B.W., N.S., I.D.L., and S.A. Y.W. and F.G. performed the experiments, calibrations, and analysed the data together with P.O.S. The numerical simulations were performed by Y.W., N.L., and P.O.S.; the analytical model was developed by N.L. and K.H. The experiment was conceived and supervised by P.O.S. All authors contributed to the discussion and interpretation of the results, and the writing of the manuscript.

% \bibliographystyle{bibstyle}
% \bibliography{all}  

\end{document}

\title{Supplementary Information for ``Precision spectroscopy by photon-recoil
 signal amplification''}
 \author{Yong Wan}
 \thanks{These authors contributed equally.}
  \affiliation{QUEST  Institut, Physikalisch-Technische Bundesanstalt, 38116 Braunschweig, Germany}
 \homepage[]{www.quantummetrology.de/quest/eqm}
 \author{Florian Gebert}
 \thanks{These authors contributed equally.}
 \affiliation{QUEST  Institut, Physikalisch-Technische Bundesanstalt, 38116 Braunschweig, Germany}
 \author{Jannes B. Wübbena}
 \affiliation{QUEST  Institut, Physikalisch-Technische Bundesanstalt, 38116 Braunschweig, Germany}
 \author{Nils Scharnhorst}
 \affiliation{QUEST  Institut, Physikalisch-Technische Bundesanstalt, 38116 Braunschweig, Germany}
 \author{Sana Amairi}
 \affiliation{QUEST  Institut, Physikalisch-Technische Bundesanstalt, 38116 Braunschweig, Germany}
 \author{Ian D. Leroux}
 \affiliation{QUEST  Institut, Physikalisch-Technische Bundesanstalt, 38116 Braunschweig, Germany}
 \author{Börge Hemmerling}
 \thanks{present address:  Department of Physics, Harvard University, Cambridge, Massachusetts 02138, USA}
 \affiliation{QUEST  Institut, Physikalisch-Technische Bundesanstalt, 38116 Braunschweig, Germany}
 \author{Niels Lörch}
 \affiliation{Institut für Theoretische Physik und Institut für Gravitationsphysik (Albert Einstein Institut), Leibniz Universität Hannover, Germany}
 \author{Klemens Hammerer}
 \affiliation{Institut für Theoretische Physik und Institut für Gravitationsphysik (Albert Einstein Institut), Leibniz Universität Hannover, Germany}
 \author{Piet O.Schmidt}
 \email{Piet.Schmidt@quantummetrology.de}
 \affiliation{QUEST  Institut, Physikalisch-Technische Bundesanstalt, 38116 Braunschweig, Germany}
 \homepage[]{www.quantummetrology.de/quest/eqm}
 \affiliation{Institut für Quantenoptik, Leibniz Universität Hannover, 30167 Hannover, Germany}

 \date{\today}
\begin{abstract}
The supplementary information provided in this document is a summary of a more detailed account published elsewhere.
\end{abstract}

 \maketitle

\section*{\Large{Supplementary Note 1}}
\textbf{Systematic shifts}

There are several effects that can introduce systematic shifts to the investigated transition.

Lineshape asymmetries can be introduced through an excitation imbalance of Zeeman-shifted line components arising from imbalanced $\sigma^+/\sigma^-$ polarization of the nominally linearly polarized spectroscopy and repump lasers.
Such an asymmetry would result in a magnetic field dependent shift of the line center. We have measured the line center for three different magnetic fields (0.387(2)~mT,~0.584(1)~mT and 0.779(3)~mT) using the two-point sampling technique. A linear fit gives a shift of -8~kHz for a magnetic field of 0.584~mT, at which the frequency measurements where performed. The uncertainty of 59~kHz is given by the $68.3\%$ prediction bound of the fit, compatible with no shift. This uncertainty is limited by the number of measurements performed.

AC Stark shifts of the transition frequency arise from differential off-resonant coupling of electro-magnetic radiation to the two involved spectroscopy states. We have investigated shifts arising from the spectroscopy laser by varying its power while simultaneously adjusting the number of spectroscopy pulses to keep the excitation signal similar.
From the intensity calibration curves similar to the ones shown in Fig.~3b of the main text, the corresponding Rabi frequencies of the spectroscopy laser have been determined to be $\Omega = 2\pi\times (4.3, 5.6,  7.3)$~MHz.
The center frequencies for different Rabi frequencies are evaluated using the two-point sampling technique.  A linear fit to the center frequency gives a shift of 60~kHz (with an uncertainty of 44~kHz given by the $68.3\%$ prediction bound of the fit) for the Rabi frequency of $\Omega=2\pi\times 5.6$~MHz chosen for the frequency measurements.
Possible ac Stark shifts from the repump laser pulse are eliminated by applying it after the spectroscopy laser pulse.
Additional ac Stark shifts can arise from the trapping rf field if the ion is not located at the node of the radial electric quadrupole. Using a two-ion crystal in a linear Paul trap, we have minimized this effect through micromotion compensation \cite{berkeland_minimization_1998} to an estimated shift of less than 1~kHz.

For the frequency measurements, we use 125~ns short spectroscopy pulses that are generated by applying rf pulses centered around 413~MHz to an acousto-optic modulator (AOM). The pulse bandwidth of $\sim 8$~MHz together with the limited spectral bandwidth of the AOM causes a shift of the envelope of the train of spectroscopy pulses, if the center frequencies are not matched or the AOM transfer function is asymmetric with respect to its center frequency.
We measure this shift for each experimental run to within 15~kHz by performing a beat measurement between the light incident onto the AOM and the first diffraction order used for spectroscopy with a fast calibrated photodetector. A similar shift arising from the photodiode is calibrated to be 22(3)~kHz. The measured shifts are applied to each frequency measurement.

Detuning-dependent Doppler shifts give rise to an asymmetry of the excitation line profile. Since we are starting in the motional ground state and the excitation signal is proportional to the ion's motion in this mode, we can accurately model this shift by either using an analytical model, or numerical simulations of the full master equation. Both are described in more detail in the next sections. The main result is shown in Supplementary Fig.~\ref{fig:dopplershift}, where we compare the shifts predicted by two analytical models and the numerical simulation implementing the full 8-level atomic system. From the numerical simulations and a conservatively estimated fluctuation of the detection signal of $\pm 0.1$, we account for a shift of $151(20)$~kHz.

\begin{figure}
  \includegraphics[width=7cm]{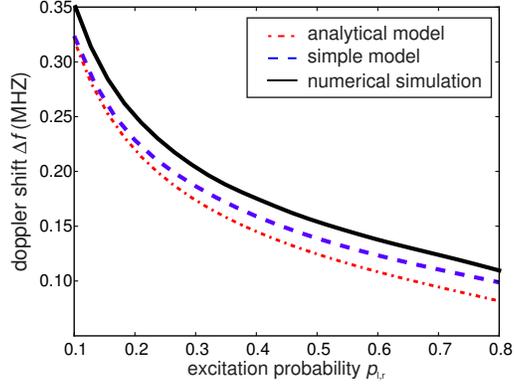}\\
  \caption{\label{fig:dopplershift}Probe frequency dependent Doppler shift versus mean excitation probability $p_\mathrm{l,r}$: Black solid line shows the shift as determined by two point sampling from full numerical solution to master equation \eqref{eq:meq}, including 8 atomic levels using the parameters of the absolute frequency measurements. Blue dashed line shows the approximate formula \eqref{eq:shift} from the 1D model neglecting diffusion. Red dash-dotted line shows the result from the analytical formula in Eq.~\eqref{eq:Ps}. The analytical formulas are correct up to second order in the Lamb-Dicke factor, which is consistent with the discrepancy from the numerical solution.}
\end{figure}

\section*{\Large{Supplementary Note 2}}
\textbf{Probe frequency dependent Doppler shift}

We derive the probe frequency dependent Doppler shift on the basis of the following simple model. We treat the $^2S_{1/2}\leftrightarrow \phantom{}^2P_{1/2}$ transition in the spectroscopy ion as a two level transition with linewidth $\Gamma$, and we restrict the description to the selected normal mode of the two-ion crystal. During each pulse the system evolves according to
  \begin{align}\label{eq:meq}
    \dot\rho&=-i\left[\omega_\mathrm{m} a^\dagger a-\frac{\Delta}{2}\sigma_z+\frac{\Omega}{2}\left(\sigma^+ e^{i\eta(a+a^\dagger)}+\mathrm{h.c.}\right),\rho\right]
    +\frac{\Gamma}{2}\mathcal{D}[\sigma^-]\rho,
  \end{align}
where $\mathcal{D}[\sigma^-]\rho=\left(2\sigma^-\rho\sigma^+-\sigma^+\sigma^-\rho-\rho\sigma^+\sigma^-\right)$, and $\sigma^+$, $\sigma^-$ are the atomic raising and lowering operators, respectively, $\sigma_z$ is a Pauli matrix, and $a^\dagger$ is the creation operator of the selected normal mode. In order to describe drift and diffusion of the normal mode we change to a Wigner function representation by introducing
    \[
    w(x,p)=\frac{1}{\pi}\int\mathrm{d}y\langle x+y|\rho|x-y\rangle e^{-2iyp}
    \]
such that the Wigner function for the normal mode is given by the trace over the internal degrees of freedom, $W(x,p)=\mathrm{tr}_{\el}\{w(x,p)\}$. In the Wigner function representation we can apply the replacement rules
  \begin{align*}
  e^{\pm i\eta(a+a^\dagger)}\rho=e^{\pm i\teta\hat{x}}\rho&\rightarrow e^{\pm i\teta x}\left(1\mp \frac{\teta}{2}\partial_p\right)w(x,p), &
  \rho e^{\pm i\eta(a+a^\dagger)}=\rho e^{\pm i\teta\hat{x}}&\rightarrow e^{\pm i\teta x}\left(1\pm \frac{\teta}{2}\partial_p\right)w(x,p).
  \end{align*}
where $\hat{x}=\frac{1}{\sqrt{2}}(a+a^\dagger)$ and we introduced the scaled Lamb-Dicke factor $\teta=\sqrt{2}\eta$. Note that this approximation is weaker than the usual Lamb-Dicke approximation. We can now perform second order perturbation theory in the terms of order $\teta$, and derive a recursion relation in order to describe the change in the Wigner function from the $n$-th to the $(n+1)$-th spectroscopy pulse. Turning the recursion relation into a differential equation yields a Fokker-Planck equation for the dynamics of the normal mode
  \begin{align}\label{eq:FPE}
    \partial_s W(x,p,s)=\left[-\left(\partial_x d_x+\partial_p d_p\right) +\frac{1}{2}\left(\partial_{xx} D_{xx}+\partial_{pp} D_{pp}+2\partial_{xp}D_{xp}\right)\right]W(x,p,s),
  \end{align}
where $s$ is the spectroscopy time (dimensionless in units of spectroscopy pulse length $\tau$). The drift and diffusion coefficients are
%\begin{eqnarray}
      \begin{align}
        d_x&=-\frac{\eta\Omega}{\sqrt{2}}\int_0^\tau\d t\sin(\omega_\mathrm{m} t)\langle \sigma_y(t)\rangle, \label{eq:coeffsfirst}\\
        d_p&=\frac{\eta\Omega}{\sqrt{2}}\int_0^\tau\d t\cos(\omega_\mathrm{m} t)\langle \sigma_y(t)\rangle, \\
				D_{xx}&=(\eta\Omega)^2\int_0^\tau\d t\int_0^t\d t' \sin(\omega_\mathrm{m} t)\sin(\omega_\mathrm{m} t')\mathrm{Re}\left(\langle \sigma_y(t)\sigma_y(t')\rangle\right)-d_x^2,\\
				D_{pp}&=(\eta\Omega)^2\int_0^\tau\d t\int_0^t\d t' \cos(\omega_\mathrm{m} t)\cos(\omega_\mathrm{m} t')\mathrm{Re}\left(\langle \sigma_y(t)\sigma_y(t')\rangle\right)-d_p^2,\\
        D_{xp}&=-(\eta\Omega)^2\int_0^\tau\d t\int_0^t\d t' \frac{1}{2}\left[\sin(\omega_\mathrm{m} t)\cos(\omega_\mathrm{m} t')+\sin(\omega_\mathrm{m} t')\cos(\omega_\mathrm{m} t)\right]\mathrm{Re}\left(\langle \sigma_y(t)\sigma_y(t')\rangle\right)-d_xd_p.\label{eq:coeffslast}
      \end{align}
%\end{eqnarray}
The mean values can be evaluated from the Bloch equations,
    \begin{align*}
      \frac{\mathrm{d}}{\mathrm{d}t}\langle {\vec\sigma}(t)\rangle&=M\langle \vec{\sigma}(t)\rangle+\Gamma \vec{m}, &
      M&=\begin{pmatrix} -\Gamma/2 & \Delta_\dop & 0 \\ -\Delta_\dop & -\Gamma/2 & -\Omega \\ 0 & \Omega & -\Gamma \end{pmatrix}, &
      \vec{m}=-\begin{pmatrix} 0 \\ 0 \\ 1 \end{pmatrix},
    \end{align*}
where $\vec{\sigma}=(\sigma_x,\sigma_y,\sigma_z)$ is a vector containing the Pauli matrices, and $\Delta_\dop(p)=\Delta-\teta\omega_\mathrm{m} p$ is a Doppler shifted detuning. For the ion being in its ground state at the beginning of each spectroscopy pulse we have to use the initial condition $\langle {\vec\sigma}(0)\rangle=\vec{m}$, such that the solution to the Bloch equations is
    %\begin{align*}
    $  \langle {\vec\sigma}(t)\rangle=\left[e^{Mt}\left(1+\Gamma M^{-1}\right)-\Gamma M^{-1}\right]\vec{m}$.
    %\end{align*}
    The two-time correlation function can be evaluated using the the quantum regression theorem
    \begin{align*}
      \frac{\mathrm{d}}{\mathrm{d}t'}\langle {\vec\sigma}(t+t')\sigma_y(t)\rangle&=M\langle \vec{\sigma}(t+t')\sigma_y(t)\rangle+\langle\sigma_y(t)\rangle\Gamma \vec{m}, &
      \langle {\vec\sigma}(t)\sigma_y(t)\rangle=\begin{pmatrix}
        i\langle\sigma_z(t)\rangle \\ 1 \\ -i\langle\sigma_x(t)\rangle
      \end{pmatrix},
    \end{align*}
    where the latter has to be used as initial condition for the first equation, and we use the notation $\langle {\vec\sigma}(t+t')\sigma_y(t)\rangle_j=\langle {\sigma_j}(t+t')\sigma_y(t)\rangle$ with $j=x,y,z$. The solution is
    %\begin{align*}
    $\langle {\vec\sigma}(t+t')\sigma_y(t)\rangle=e^{Mt'}\left(\langle {\vec\sigma}(t)\sigma_y(t)\rangle+\langle\sigma_y(t)\rangle\Gamma M^{-1}\vec{m}\right)- \langle\sigma_y(t)\rangle\Gamma M^{-1}\vec{m}.$
    %\end{align*}

    The solutions for $\langle\sigma_y(t)\rangle$ and $\langle\sigma_y(t)\sigma_y(t')\rangle$ can be inserted into the expressions for the drift and diffusion coefficients. The remaining time integrals can in principle be evaluated analytically. Note that due to the Doppler shift in the detuning $\Delta_\dop$ all drift and diffusion coefficients in Eq.~\eqref{eq:FPE} have in principle a nonlinear dependence on the momentum variable $p$. However, in first order of Lamb-Dicke expansion only the drift in momentum $d_p$ is Doppler shifted such that we can approximate
      \begin{align}\label{eq:dop}
      d_p(p)&=d_p(0)+\frac{\partial d_p}{\partial \Delta_\mathrm{dop}}\frac{\partial\Delta_\mathrm{dop}}{\partial p}p
      =d_p-g p, &
      g&=\sqrt{2}\eta\omega_\mathrm{m} \frac{\partial d_p}{\partial \Delta}.
      \end{align}
    $g$ thus describes a Doppler shift induced (anti)damping rate for positive (negative) values of $g$, that is for red (blue) detuning from resonance. All other drift and diffusion coefficients show a weaker dependence on the momentum variable of higher order in the Lamb-Dicke parameter, such that we can neglect their dependence on $p$ and set $\Delta_\dop=\Delta$ when evaluating Eqs.~\eqref{eq:coeffsfirst}-\eqref{eq:coeffslast}. As a result all drift and diffusion coefficients will be even functions in the detuning, and $g$ will be an odd function.

\paragraph*{Exact solution to diffusion equation ---} With the linear approximation \eqref{eq:dop} for the Doppler shift the Fokker-Planck equation \eqref{eq:FPE} describes a Gaussian dynamics. When the normal mode is prepared initially in its ground state, the state will be fully characterized by its first and second moments at all subsequent times. These are conveniently collected in the displacement vector $\vec{r}=\langle \vec{R}\rangle$ and the covariance matrix $\gamma_{ij}=\langle R_iR_j+R_jR_i\rangle-2 r_ir_j$ where $\vec{R}=(\hat{x},\hat{p})^T$ and $i,j=1,2$. Solving the Fokker-Planck with initial conditions $\vec{r}=0$ and $\gamma=\mathds{1}$ (the $2\times 2$ identity matrix) one finds
        \begin{align}\label{eq:dispcov}
        \vec{r}(s)&=\begin{pmatrix}
          d_x s \\
          \frac{1-e^{-gs}}{g}d_p
        \end{pmatrix},  &
        \gamma(s)&=\begin{pmatrix}
          1+2D_{xx}s    & \frac{1-e^{-gs}}{g}2D_{xp}\\ \frac{1-e^{-gs}}{g}2D_{xp} & \quad e^{-2gs}+\frac{1-e^{-2gs}}{2g}2D_{pp}
        \end{pmatrix}.
        \end{align}
The probability to excite the ion from its ground state after a spectroscopy time $s$ follows as
    \begin{align}\label{eq:Ps}
      P(s)=1-\left|\frac{\mathds{1}+\gamma(s)}{2}\right|^{-1/2}\exp\left[-\vec{r}(s)^{T}(\mathds{1}+\gamma(s))^{-1}\vec{r}(s)\right],
    \end{align}
from which the shift $\Delta f$ can be extracted numerically using the two point sampling method.

%The result is compared with the crude approximation \eqref{eq:shift} and the result from a numerical solution of the master equation \eqref{eq:meq} in Fig.~\ref{fig:dopplershift}.

\paragraph*{Estimate for one dimension ---} In order to arrive at a simple physical picture of the Doppler induced shift we neglect the small drift and diffusion in position, that is, we set $d_x=D_{xx}=D_{xp}=0$. These coefficients will be smaller than the respective ones for the momentum by at least a factor $(\omega_\mathrm{m}\tau)$ (trap frequency times pulse length). Thus, we effectively reduce our description to a one dimensional drift and diffusion process for the momentum. The mean momentum $\langle \hat{p}\rangle=r_2$ and its variance $v_p=\langle p^2\rangle-\langle p\rangle^2=\gamma_{22}/2$ evolve, according to Eq.~\eqref{eq:dispcov}, as
\begin{align*}
  \langle p(s)\rangle&=\frac{1-e^{-gs}}{g}d\simeq ds-\frac{ds^2}{2}g,  &
  v_p(s)&=e^{-2gs}+\frac{1-e^{-2gs}}{g}D_{pp}\simeq 1+2(D_{pp}-g)s.
\end{align*}
The approximations represent first order Taylor expansions in the Doppler shift induced damping/heating rate $g$ and diffusion $D_{pp}$. Note that, in leading order, the Doppler shift contributes to the variance \emph{linearly} and to the mean momentum \emph{quadratically} in $s$. In this order of $s$, and for the parameter regime considered here, the diffusion $D_{pp}$ will make a negligible contribution as compared to $g$. In the following we will therefore set $D_{pp}=0$. With the current approximation the excitation probability \eqref{eq:Ps} becomes
\begin{align*}
  P(s)=1-\left[\frac{2}{1+v_p(s)}\right]^{1/2}\exp\left[-\frac{\langle p(s)\rangle^2}{1+v_p(s)}\right]\simeq P_0(s)+\delta P(s),
\end{align*}
where 
    \begin{align*}
    P_0(s)&=1-\exp\left[-\frac{d_p^2s^2}{2}\right], &
    \delta P(s)&=\exp\left[-\frac{d_p^2s^2}{2}\right]\frac{gs}{2},
    \end{align*}
are, respectively, the excitation probability without Doppler shift and the first order correction in $g$. $P_0(s)$ will be an entirely even function in the detuning, while the small correction $\delta P(s)$ is odd in the detuning. $P_0(s)$ reproduces the scaling with the number of spectroscopy pulses and the Lamb-Dicke factor as stated in the main text. For a given spectroscopy time $s$ the shift $\Delta f$ of the center of $P(s)$ away from $\Delta=0$, as determined by the two point sampling method, therefore is
    \[
    2\pi\Delta f=\frac{\delta P(s)}{\frac{\partial P_0(s)}{\partial\Delta}}=\frac{g}{2d_ps\frac{\partial d_p}{\partial\Delta}}=\frac{\eta\omega_\mathrm{m} }{\sqrt{2}d_ps},
    \]
where we used the definition of $g$ in Eq.~\eqref{eq:dop}. If the two point sampling method is applied at a fixed motional excitation probability $p_\mathrm{l,r}$ the detuning $\Delta$ (and therefore the drift per pulse $d_p$) is effectively tied to the spectroscopy time such that (to zeroth order in $g$) $p_\mathrm{l,r}=P_0(s)$ or, equivalently, $d_ps=\sqrt{-2\log (1-p_\mathrm{l,r})}$. Overall the shift, as measured in the two point sampling method with signal amplitude $p_\mathrm{l,r}$ is roughly independent from the spectroscopy time, and given by
    \begin{align}\label{eq:shift}
    2\pi\Delta f=\frac{\eta\omega_\mathrm{m} }{2}\left[\log\left(\frac{1}{1-p_\mathrm{l,r}}\right)\right]^{-1/2},
    \end{align}
as stated in the main text.

\section*{\Large{Supplementary Note 3}}
\textbf{Numerical Simulations}

Numerical simulations have been performed using QuTiP \cite{johansson_qutip_2013}. For this, the Master equation Eq.~\ref{eq:meq} is implemented in 1$^\mathrm{st}$ order Lamb-Dicke approximation. To include the effect of Zeeman broadening and optical pumping, the master equation \eqref{eq:meq} is expanded into a system containing eight electronic levels. Up to 20 motional levels of the normal mode are taken into account. The effect of recoil from spontaneous emission on the final spectroscopy signal has been found to be negligible at the 1~kHz level and has therefore been omitted. Using the experimentally determined parameters for the magnetic field, pulse length and Rabi frequency, we simulate the resonance curve as shown in Fig.~3a, which agrees well with the experimental data.
The Rabi frequency of the spectroscopy laser is found by simultaneous fitting of simulated excitation signals as a function of the number of pulses to data of the observed signal with and without repumper. Fit parameters are the Rabi frequency, a signal offset and an amplitude reduction factor arising from experimental imperfections. The signal strength as a function of the number of absorbed photons are extracted from the simulated \dpoh population of the signal with repumper. An excitation signal of 0.5 corresponding to a SNR of one is reached after 40 spectroscopy/repump pulses. The integrated \dpoh-state population corresponds to 9.8(1.2) absorbed photons.
From simulated resonances curves similar to Fig.~3a using the experimentally calibrated Rabi frequency $\Omega = 2\pi\times 5.6$~MHz and 70 excitation pulses with a pulse length of $\tau=125$~ns, a lineshape shift of 151(20)~kHz can be extracted for the two-point sampling technique, assuming conservatively estimated fluctuations of the detection signal of $\pm 0.1$.

%\bibliographystyle{bibstyle}
%\bibliography{all}